\documentclass[twoside,journey]{IEEEtran}
\usepackage{makecell}
\usepackage{hyperref}
\usepackage{array}
\usepackage{graphicx,amssymb,amsmath}
\usepackage{multicol}
\usepackage[noadjust]{cite}
\usepackage{setspace}
\usepackage{subcaption}
\usepackage{graphicx}
\usepackage{float}
\usepackage {url}
\usepackage{stfloats}
\usepackage{amsthm,pifont}
\usepackage{flushend}
\usepackage{cases,subeqnarray}
\usepackage{bm,multirow,bigstrut}
\usepackage{amsmath, amsthm, amssymb}
\usepackage{textcomp}
\usepackage{latexsym,bm}
\usepackage{booktabs}
\usepackage{xcolor}
\usepackage{mathtools}
\usepackage{dsfont}
\usepackage{extarrows}
\usepackage{epsfig}
\usepackage{epsfig}
\usepackage{epstopdf}
\usepackage[noend]{algpseudocode}
\usepackage{algorithmicx,algorithm}
\usepackage{makecell}
\usepackage{colortbl}
\usepackage{booktabs}
\usepackage{graphicx} 
\usepackage{array}    
\usepackage{hyperref} 
\usepackage{booktabs}
\usepackage{enumitem}
\usepackage{diagbox}
\usepackage{tikz}
\usetikzlibrary{trees,shadows.blur}
\usepackage{forest}

\usepackage[table]{xcolor}
\usepackage{colortbl}
\usepackage{booktabs}
\usepackage{tabularx}
\usepackage{makecell}
\usepackage{array}

\definecolor{HeaderBlue}{RGB}{78,129,171}   
\definecolor{GroupOrange}{RGB}{240,165,70}  
\definecolor{LightGray}{RGB}{245,245,245}

\usepackage[table]{xcolor}
\usepackage{colortbl}

\definecolor{myorange}{RGB}{255,230,204} 
\definecolor{myblue}{RGB}{214,232,255}   
\definecolor{myheader}{RGB}{235,235,235} 

\theoremstyle{plain}

\theoremstyle{plain}

\usepackage{amsmath}

\newcommand{\ignore}[1]{{{\color{yellow} }}}
\renewcommand{\arraystretch}{1.5} 
\definecolor{blue-green}{rgb}{0.0, 0.87, 0.87}
\IEEEoverridecommandlockouts
\begin{document}


\title{Wireless Context Engineering for Efficient Mobile Agentic AI and Edge General Intelligence}
\author{Changyuan Zhao, Jiacheng Wang, Yunting Xu, Zan Li,~\IEEEmembership{Fellow,~IEEE},\\ Abbas Jamalipour,~\IEEEmembership{Fellow,~IEEE}, Dong In Kim,~\IEEEmembership{Life Fellow,~IEEE}
    \thanks{C. Zhao, J.~Wang and Y.~Xu, are with the College of Computing and Data Science, Nanyang Technological University, Singapore (e-mail: zhao0441@e.ntu.edu.sg, jiacheng.wang@ntu.edu.sg, yuntingxu622@gmail.com).}
        \thanks{Z. Li is with the State Key
Laboratory of Integrated Services Networks, Xidian University, Xian 710071,
China (e-mail: zanli@xidian.edu.cn).}
\thanks{A. Jamalipour is with the School of Electrical and Computer Engineering, University of Sydney, Australia, and with the Graduate School of Information Sciences, Tohoku University, Japan (e-mail: a.jamalipour@ieee.org).}
    \thanks{D. I. Kim is with the Department of Electrical and Computer Engineering, Sungkyunkwan University, Suwon 16419, South Korea (e-mail: dongin@skku.edu).}}

\maketitle
\vspace{-1cm}

\begin{abstract}
Future wireless networks demand increasingly powerful intelligence to support sensing, communication, and autonomous decision-making. While scaling laws suggest improving performance by enlarging model capacity, practical edge deployments are fundamentally constrained by latency, energy, and memory, making unlimited model scaling infeasible. This creates a critical need to maximize the utility of limited inference-time inputs by filtering redundant observations and focusing on high-impact data. In large language models and generative artificial intelligence (AI), context engineering has emerged as a key paradigm to guide inference by selectively structuring and injecting task-relevant information. Inspired by this success, we extend context engineering to wireless systems, providing a systematic way to enhance edge AI performance without increasing model complexity. In dynamic environments, for example, beam prediction can benefit from augmenting instantaneous channel measurements with contextual cues such as user mobility trends or environment-aware propagation priors. We formally introduce wireless context engineering and propose a Wireless Context Communication Framework (WCCF) to adaptively orchestrate wireless context under inference-time constraints. This work provides researchers with a foundational perspective and practical design dimensions to manage the wireless context of wireless edge intelligence. An ISAC-enabled beam prediction case study illustrates the effectiveness of the proposed paradigm under constrained sensing budgets.
\end{abstract}
\IEEEpeerreviewmaketitle

\section{Introduction}\label{intro}


Artificial intelligence (AI) systems fundamentally rely on input information to perform inference and decision-making. However, a fundamental challenge persists: \textit{the amount of information that an AI system can effectively process at inference time remains inherently limited}~\cite{liu2024lost}.
Over the past decade, scaling laws have driven a dominant response to this limitation by motivating increasingly large models with expanded parameter capacity. This paradigm has given rise to powerful architectures, including large language models (LLMs), diffusion models, and world models, which exhibit enhanced representation power and generalization ability~\cite{zhao2025agentification}.
In parallel, a larger model scale has enabled the expansion of inference-time context windows, but this gain comes at a substantial cost. For example, in the Qwen family, model size increases from 7B/14B to tens of billions of parameters, while the context window expands to at most 128K tokens, representing a limited gain relative to model scaling\footnote{https://qwenlm.github.io/blog/qwen3}. Moreover, due to attention and Key-Value (KV)-cache usage, inference latency and memory consumption grow superlinearly with context length, making long-context inference increasingly time- and energy-intensive~\cite{mei2025survey}.
Consequently, as AI systems evolve toward edge-oriented and agentic intelligence, model scale and context window size can no longer be adjusted freely. Practical deployments are constrained by computation, memory, and energy budgets, shifting the central challenge from increasing model capacity to effectively utilizing limited input information at inference time.


To illustrate this shift in focus, consider a wireless network operating in a dynamic environment. At any given moment, the system may have access to a wide range of information, including instantaneous channel measurements, user mobility patterns, traffic demand, service requirements, historical control actions, and environmental conditions. All of these factors could influence scheduling, resource allocation, or control decisions.
However, it is neither feasible nor beneficial to incorporate all available information into a fixed AI model at inference time, as doing so would overwhelm the inference compute budget and introduce noise.
For instance, a channel log collected in a static environment may become irrelevant or incorrect when a user suddenly accelerates.


We define this selected, retained information as \textbf{context}. Distinct from a momentary snapshot, such as a single channel state information (CSI) pilot, \textit{context represents the structured set of information, spanning past history, current state, external knowledge, and future intent, that conditions the AI's inference}~\cite{mei2025survey}. Just as a well-crafted prompt guides an LLM to the correct answer without retraining the model, a well-engineered wireless context guides an edge agent to an optimal decision.
However, managing this context in wireless systems is particularly challenging due to its heterogeneity and multi-timescale characteristics. Different contextual elements evolve at disparate time scales, such as channel conditions fluctuating on the order of milliseconds while traffic dynamics change over much longer periods. Without active management, stale or redundant information may accumulate in the context, potentially distorting inference and degrading decision quality.
This motivates a runtime view in which context is treated as a 
bounded system resource rather than an unlimited input stream.

Motivated by the above challenges, we introduce wireless context engineering as a system-level and inference-time approach for governing the handling of contextual information in wireless edge intelligence systems.
\textit{Wireless context engineering focuses on selecting, structuring, compressing, updating, and delivering heterogeneous wireless context, such as channel conditions, mobility patterns, traffic dynamics, service requirements, and environmental factors, so that limited inference-time context capacity is used efficiently.} Rather than modifying the underlying AI model, its objective is to extract and organize core, task-relevant information from raw and evolving wireless observations into a compact and structured form, akin to meticulously designed knowledge, thereby improving inference and decision-making performance at the edge~\cite{zhang2025agentic}.

From a system perspective, wireless context engineering determines how raw wireless signals are transformed into agent-interpretable and structured input, and how long such information remains valid, which is particularly important for modern learning paradigms, including wireless foundation models, that exploit large-scale and multi-modal wireless data~\cite{xu2024large}. Rather than modifying the underlying model, context engineering governs the quality and temporal relevance of information exposed at inference time, thereby directly shaping key aspects of system performance:
\begin{itemize}
    \item \textbf{Robustness:} 
    By designing uncertainty-aware context, such as long-term channel statistics, mobility trends, and interference regimes, wireless context engineering conditions generation and decision-making on persistent wireless states rather than instantaneous CSI, thereby suppressing transient fluctuations and improving robustness under non-stationary channels.

    \item \textbf{Efficiency:} 
    By encoding task-relevant compact context, including task intent, semantic importance, and service constraints, wireless context engineering aligns the generation process with decision objectives, enabling high-fidelity signal generation with reduced data transmission and inference overhead.

    \item \textbf{Coherence:} 
    By maintaining historical and predictive context, such as past actions, trajectory states, and long-term objectives, wireless context engineering explicitly couples current decisions with future system evolution, ensuring temporally consistent resource allocation and control.
\end{itemize}

To this end, this paper presents a comprehensive overview of wireless context engineering as a foundational perspective for enabling effective agentic intelligence in wireless edge systems.
The key contributions of this work are summarized as follows:
\begin{itemize}
    \item We systematically define wireless context by clarifying its concept, sources, and scope across multiple layers of wireless systems. We further explain how wireless context differs from instantaneous observations and how it shapes the behavior of modern AI models under limited inference-time input capacity.
    \item We identify key design dimensions of wireless context engineering, including acquisition, structuring, compression and prioritization, persistence and aging, and delivery and access. We then introduce a set of performance metrics to evaluate wireless context engineering.
    \item We propose a Wireless Context Communication Framework (WCCF) to engineer adaptive multimodal context under finite bandwidth and inference-time constraints. An integrated sensing and communication (ISAC)–enabled vehicle-to-infrastructure (V2I) beam prediction case study shows that dynamic context selection achieves an optimized accuracy–cost trade-off.
\end{itemize}

We expect this work to serve as a springboard for future developments in wireless context engineering for edge intelligence.
With this paper, researchers will be able to
\begin{itemize}
\item Gain an understanding of wireless context, including its definition, sources, and role in wireless intelligence.

\item Learn how to design and evaluate wireless context engineering under limited inference-time input capacity.

\item Understand how context communication frameworks can be used to enhance wireless task performance at the edge.
\end{itemize}

\section{Overview of Wireless Context Engineering}

This section introduces the concept of wireless context in agentic wireless systems, explains how it shapes agent behavior across representative wireless tasks, and distills wireless context engineering into a set of practical design dimensions.

\subsection{Wireless Context: Concept, Scope, and Characteristics}

In agentic wireless systems, \textit{wireless context refers to the information that conditions the perception, reasoning, and action of intelligent agents beyond instantaneous signal observations.} Unlike raw measurements such as a single snapshot of CSI, wireless context captures situational knowledge that enables agents to interpret observations, anticipate system evolution, and select actions over time. From this perspective, the information exposed to an agent can be viewed as comprising \textbf{task information }and \textbf{contextual information}. Task information represents the minimal, decision-critical inputs required to execute an action, whereas contextual information provides auxiliary cues that help refine or disambiguate that decision.

This distinction can be understood through navigation. 
A task instruction, such as ``turn left'' is insufficient without 
nearby landmarks, but excessive landmark descriptions may 
obscure the actual decision. Similarly, wireless agents need 
compact contextual cues rather than all available observations.

A key characteristic of wireless context is its heterogeneity across multiple system layers and modalities. In practical wireless systems, contextual information may arise from diverse sources, including but not limited to:
\begin{itemize}
    \item \textbf{Physical-layer context:} long-term channel statistics, interference, sensing uncertainty, and signal quality.
    \item \textbf{Network-layer context:} queue backlogs, traffic load variations, topology changes, and resource availability.
    \item \textbf{Service- and task-level context:} latency constraints, reliability targets, semantic importance of transmitted information, and task priorities.
    \item \textbf{Environmental context:} user mobility patterns, obstacles, propagation environments, and external conditions that influence wireless links.
\end{itemize}


\begin{figure*}[htp]
    \centering
    \includegraphics[width= 0.90\linewidth]{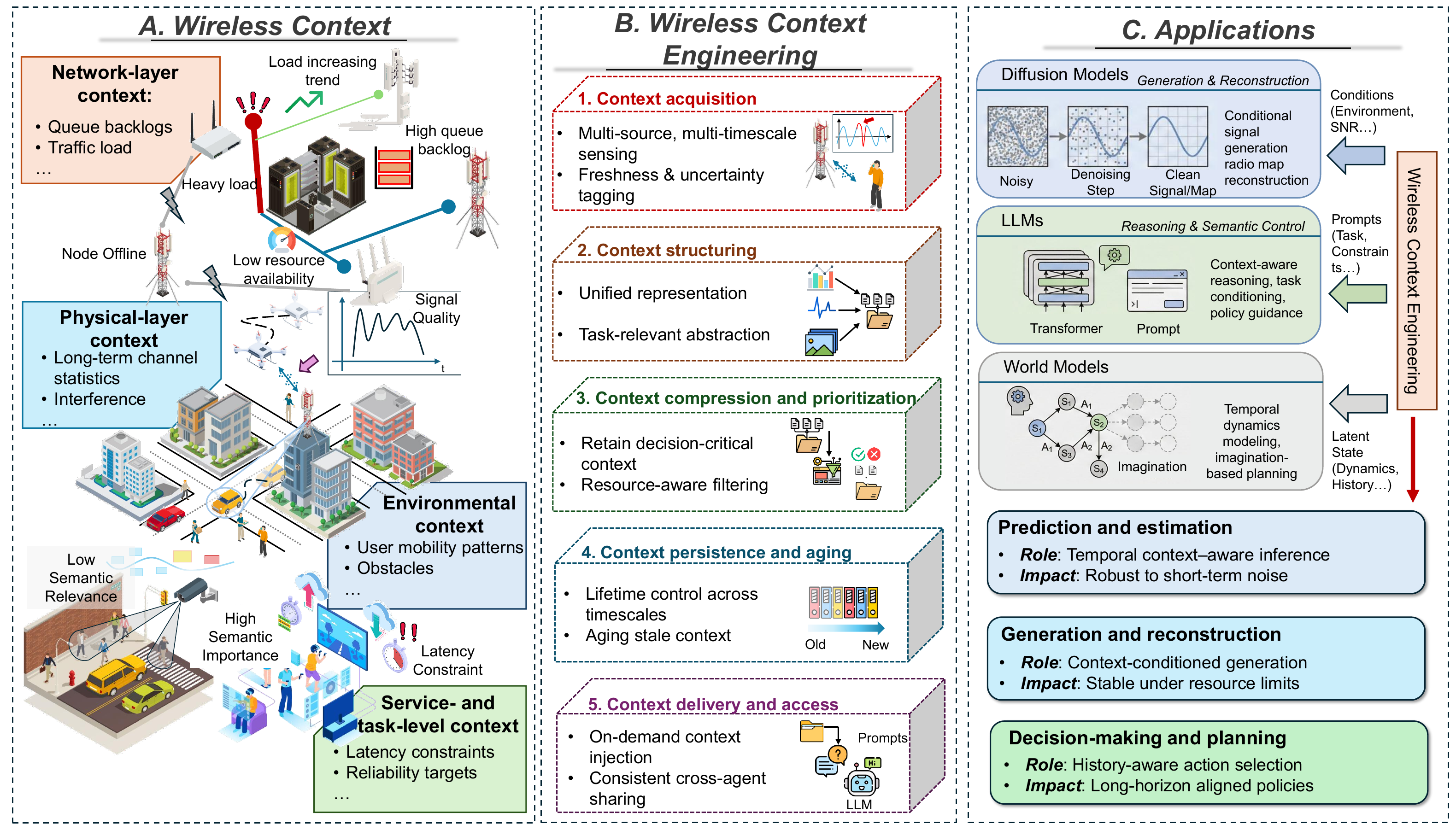}
    \caption{Illustration of wireless context engineering for agentic intelligence in wireless networks.
Part A shows multi-layer wireless context spanning physical, network, environmental, and service/task levels.
Part B illustrates wireless context engineering across five dimensions, including acquisition, structuring, compression, persistence, and delivery.
Part C presents context-conditioned prediction, generation, and decision-making enabled by agentic models in dynamic wireless environments.}
    \label{fig:frame}
\end{figure*}

\subsection{Wireless Context in Intelligent Wireless Applications}

Wireless context is not merely auxiliary data. It fundamentally reshapes how agentic intelligence interacts with the physical world. This influence is distinct across three core wireless task categories, where context acts as a bridge between limited onboard resources and complex environmental dynamics.
\begin{itemize}
    \item \textbf{Prediction and estimation:} 
    In wireless prediction tasks, such as channel state estimation and interference forecasting, wireless context enables agents to reason over temporally extended information rather than relying solely on instantaneous measurements, thereby stabilizing inference under noisy and partially observable conditions.
    For example, recent work~\cite{xu2025enhancing} shows that augmenting radio frequency (RF)-based predictors with environment-aware context extracted from pretrained large vision models (LVMs) significantly improves beamforming and localization performance compared to RF-only approaches, illustrating how context knowledge helps distinguish persistent wireless conditions from short-term fluctuations.
    From a model perspective, such context is typically incorporated as augmented inputs or latent belief states, allowing predictors to operate effectively under limited inference-time input capacity~\cite{zhao2025edge}.

    \item \textbf{Generation and reconstruction:}
    For generation-oriented tasks such as semantic communication and signal reconstruction, wireless context constrains feasible outputs and guides generation toward task-consistent results. By incorporating task intent, communication states, and environmental conditions, context enables efficient generation under limited communication and computation resources~\cite{liu2025context}. For example, CaSemCom employs an LLM-based gating mechanism to adaptively select high-impact semantic features based on task and channel context, substantially improving reconstruction fidelity and bandwidth efficiency over static baselines~\cite{liu2025context}. This context-conditioned generation preserves semantic relevance while minimizing transmission overhead.

    
    \item \textbf{Decision-making and planning:} 
    In strategy-oriented tasks such as resource allocation, scheduling, and trajectory planning, wireless context enables agents to align short-term actions with long-term objectives by reasoning over historical and predictive information rather than instantaneous observations, leading to more stable and coordinated system behavior~\cite{xu2025integrating}. For example, world model-based approaches such as Wireless Dreamer~\cite{zhao2025world} encode predictive context through learned latent dynamics and imagined future trajectories. This predictive context enables more effective long-horizon UAV planning than purely reactive, model-free policies. In decision-making systems, such context is typically incorporated into abstract state representations or memory structures, allowing policies to operate without full system observability at each decision step.
    

\end{itemize}

Across these domains, the unifying insight is that wireless context engineering is not about maximizing information quantity, but optimizing information density under inference constraints. Directly exposing edge models to raw, high-dimensional history often overwhelms their limited context windows, leading to latency spikes or decision bias. Instead, well-engineered context functions as an efficiency filter, compressing vast raw observations into compact, knowledge-rich inputs. This allows edge agents to emulate the performance of larger, distinct-state models while operating within the tight computation and memory budgets of the wireless edge.

\subsection{Design of Wireless Context Engineering}

Unlike conventional optimization pipelines that operate on fixed numerical matrices, agentic AI reasons over tokens, prompts, or latent states. 
This shift motivates the need for wireless context engineering not merely as data processing, but as a distinct knowledge discovery process.
It focuses on how raw, high-dimensional wireless context are distilled into knowledge-rich representations, such as semantic prompts or structured memory, and how such context is governed at runtime~\cite{mei2025survey}.


To bridge the gap between complex wireless environments and the limited inference capabilities of edge agents, wireless context engineering can be characterized through five key design dimensions. We illustrate these dimensions using a running example of a UAV agent performing path planning in a wireless network.


\begin{itemize}
    \item \textbf{Context acquisition:}
    Refers to how contextual information is obtained from wireless systems and the environment. Agents actively decide which tools to invoke and which historical information to retrieve.
    
    \emph{Example:} Instead of passively receiving all data, a UAV agent must decide whether to incur the latency cost of triggering a real-time CSI pilot sequence or to simply query a historical radio map from the database~\cite{zeng2024tutorial}.
    Under limited model input capacity, these acquisition decisions determine which signals are exposed to the model.

    \item \textbf{Context structuring:}
    Describes how acquired context is organized and represented before being consumed~\cite{zhang2025agentic}. This dimension governs how heterogeneous wireless information is formatted into compact, model-interpretable inputs rather than raw signal streams.

    \emph{Example:} 
    In high-mobility UAV scenarios, relying solely on transient pilots is inefficient. Instead, context structuring fuses sparse real-time observations with a pre-learned Channel Knowledge Map (CKM)~\cite{zeng2024tutorial}. By organizing location-specific priors, such as blockage probabilities and dominant angles, into a structured representation, this approach minimizes pilot overhead while enhancing channel prediction robustness against dynamic variations.

    

    \item \textbf{Context compression and prioritization:}
Determines how contextual information is summarized, reduced, or ranked. This process prioritizes decision-critical features to maximize information density, transforming high-dimensional raw data into compact representations while suppressing irrelevant noise.

    
    \emph{Example:} Consider a UAV performing integrated sensing and communication (ISAC) beam tracking. Since streaming raw radar and CSI data would saturate the limited context window, a world model encoder compresses them into compact latent states, retaining only alignment-critical geometric and channel features~\cite{zhao2025edge}. This efficient representation prevents window overflow, enabling the agent to leverage long-term context for robust blockage prediction.

    

    \item \textbf{Context persistence and aging:}
    Captures how long contextual information remains valid and how it evolves~\cite{zhang2025agentic}. This governs how context is stored in agent memory, determining which historical states and stable descriptors are retained, refreshed, or discarded.
    
    \emph{Example:} The system maintains persistent context for static environmental features, such as radio maps, while rapidly aging out transient context, including millisecond-level channel fluctuations~\cite{zeng2024tutorial}.

    \item \textbf{Context delivery and access:}
    Describes how structured wireless context is delivered to intelligent components. This regulates the timing and scope of context injection, ensuring prompts, retrieved data, and tool outputs are provided exactly when needed~\cite{mei2025survey}.
    
    \emph{Example:} Instead of flooding the agent with continuous updates, context is delivered proactively only when the UAV approaches a decision point, such as a trajectory intersection, ensuring the model is not overwhelmed.
\end{itemize}


In summary, agent behavior is strongly influenced by how wireless context is handled at runtime. By explicitly designing these five dimensions, wireless systems can achieve stable, efficient, and goal-aligned intelligence without increasing model complexity, serving as a practical lever for improving edge intelligence under real-world constraints.

\subsection{Wireless Context Engineering and Related Paradigms}
\label{subsec:wce_vs_existing}

Wireless context engineering is closely related to several existing paradigms, including context-aware networking, multimodal fusion, and semantic communication. However, as summarized in Table~\ref{tab:wce_vs_existing}, these paradigms primarily focus on optimization, representation integration, or semantic transmission, whereas wireless context engineering focuses on governing what information should be acquired and exposed to resource-constrained AI agents at inference time~\cite{mei2025survey}.

Rather than replacing existing paradigms, wireless context engineering complements them by managing the lifecycle of context under inference-time constraints~\cite{zhang2025agentic}. For example, multimodal fusion mainly focuses on how different modalities are integrated after acquisition, whereas wireless context engineering also determines whether these modalities should be acquired in the first place. In a beam prediction task, a fusion-based system may continuously use GPS, camera, and LiDAR features whenever they are available. Such a fixed strategy can waste bandwidth and computation in stable line-of-sight conditions, or fail to identify when high-cost visual and LiDAR context is necessary under blockage. Wireless context engineering addresses this issue by treating context as a bounded runtime resource: lightweight location context can be used when the beam direction is stable, whereas additional visual or geometric context is retrieved only when it is expected to improve the decision. Similarly, semantic communication focuses on what information should be transmitted over the channel, while wireless context engineering further governs how contextual information is selected, updated, aged, and injected into the agent's inference-time context. 

Therefore, its key distinction lies in context governance before, during, and after inference, rather than in representation fusion or semantic transmission alone.

\begin{table*}[t]
\centering
\scriptsize
\caption{Wireless context engineering in relation to existing paradigms.}
\label{tab:wce_vs_existing}
\setlength{\tabcolsep}{2.3pt}
\renewcommand{\arraystretch}{1.03}

\begin{tabular}{
>{\raggedright\arraybackslash}p{2.0cm}
>{\raggedright\arraybackslash}p{2.9cm}
>{\raggedright\arraybackslash}p{4.2cm}
>{\raggedright\arraybackslash}p{2.9cm}
>{\raggedright\arraybackslash}p{3.5cm}}
\hline
\rowcolor{myheader}
Paradigm & Primary objective & Key characteristics & Wireless usage & Key takeaway \\
\hline

\cellcolor{myorange}Context-aware networking
&
\parbox[t]{\linewidth}{\raggedright
\textbullet\;Improve optimization using auxiliary information
}
&
\parbox[t]{\linewidth}{\raggedright
\textbullet\;Context appended to model inputs\\
\textbullet\;Supports downstream decisions\\
\textbullet\;Focus on feature-level utilization
}
&
\parbox[t]{\linewidth}{\raggedright
\textbullet\;Scheduling and resource allocation\\
\textbullet\;Mobility and traffic awareness
}
&
\cellcolor{myblue}\parbox[t]{\linewidth}{\raggedright
Context acts as auxiliary side information for optimization
}
\\
\hline

\cellcolor{myorange}Multimodal fusion
&
\parbox[t]{\linewidth}{\raggedright
\textbullet\;Learn unified multimodal representations
}
&
\parbox[t]{\linewidth}{\raggedright
\textbullet\;Fuse wireless and sensory features\\
\textbullet\;Focus on representation integration after modality acquisition\\
\textbullet\;Improve robustness through complementary modalities
}
&
\parbox[t]{\linewidth}{\raggedright
\textbullet\;Beam prediction and localization\\
\textbullet\;Environment-aware sensing
}
&
\cellcolor{myblue}\parbox[t]{\linewidth}{\raggedright
Focuses on how modalities are fused after acquisition
}
\\
\hline

\cellcolor{myorange}Semantic communication
&
\parbox[t]{\linewidth}{\raggedright
\textbullet\;Transmit task-relevant semantics efficiently
}
&
\parbox[t]{\linewidth}{\raggedright
\textbullet\;Encode semantic information instead of raw bits\\
\textbullet\;Reduce communication overhead\\
\textbullet\;Focus on transmission and reconstruction
}
&
\parbox[t]{\linewidth}{\raggedright
\textbullet\;Semantic transmission\\
\textbullet\;Task-oriented communication
}
&
\cellcolor{myblue}\parbox[t]{\linewidth}{\raggedright
Focuses on what information should be transmitted over the channel
}
\\
\hline

\cellcolor{myorange}Wireless context engineering
&
\parbox[t]{\linewidth}{\raggedright
\textbullet\;Govern inference-time context exposure
}
&
\parbox[t]{\linewidth}{\raggedright
\textbullet\;Treat context as a runtime resource\\
\textbullet\;Selectively acquire and structure context\\
\textbullet\;Manage persistence and delivery under bounded inference-time capacity
}
&
\parbox[t]{\linewidth}{\raggedright
\textbullet\;Agentic wireless intelligence\\
\textbullet\;Multimodal beam prediction\\
\textbullet\;Multi-timescale context management
}
&
\cellcolor{myblue}\parbox[t]{\linewidth}{\raggedright
Govern what context should be acquired and exposed at inference time
}
\\
\hline

\end{tabular}
\end{table*}

\subsection{Performance Metrics for Wireless Context Engineering}

The performance of wireless context engineering can be evaluated from four complementary dimensions that jointly reflect how efficiently contextual information is utilized by agentic models and how effectively it translates into wireless system gains~\cite{hu2024unveiling}.
\textbf{Information efficiency} focuses on how effectively limited inference-time context capacity is utilized, including perplexity for context-conditioned inference, information density for decision-relevant context, and context window utilization for inference-time memory.
\textbf{Agentic execution efficiency} characterizes how context improves action effectiveness, measured by task success rate for quality of service (QoS) satisfaction, convergence time for control and scheduling, and tool usage accuracy for resource allocation.
\textbf{Fidelity and robustness} assess the reliability of context-conditioned inference, including hallucination rate for physical consistency and needle-in-a-haystack accuracy for critical event detection.
\textbf{System cost} captures the overhead of context handling, measured by time to first token for inference latency and KV cache size for memory footprint.

\vspace{-0.5em}
\section{Wireless Context Engineering Techniques}





\subsection{Context Selection}

Context selection determines which subset of wireless context is exposed to an AI model, given limited inference-time input capacity.
In vision-aided wireless systems, directly using raw RGB images can introduce substantial background redundancy, obscuring propagation-relevant features and degrading prediction accuracy.
In~\cite{zhang2025vision}, the authors propose a two-stage vision-aided channel prediction framework that first extracts selective visual context before channel inference.
Specifically, object detection and instance segmentation are employed to isolate task-relevant visual entities, such as the target vehicle and dominant scatterers, while suppressing irrelevant background information.
Experimental results demonstrate that selecting appropriate visual context significantly improves prediction fidelity, with instance-segmentation-based context reducing received-power prediction RMSE by approximately 15\% to 25\% compared with bounding-box-based inputs, and by over 40\% compared with binary-mask representations across multiple vehicular scenarios~\cite{zhang2025vision}. This result validates the effectiveness of context selection in vision-assisted channel prediction.

\vspace{-1em}
\subsection{Context Compression}

Context compression focuses on reducing the volume of wireless context while preserving task-relevant information under stringent bandwidth constraints.
In immersive 6G scenarios, directly transmitting raw multimodal context incurs prohibitive bandwidth overhead, making semantic-level compression indispensable.
In \cite{zhang2025multimodal}, the authors propose an MLLM-integrated semantic communication framework that performs importance-aware context compression guided by multimodal LLMs.
By generating semantic attention maps that identify critical content regions, the importance-aware encoder allocates higher transmission fidelity to semantically important context while aggressively compressing non-critical information.
Experimental results show that under an ultra-low compression ratio of 1.3\%, the proposed approach improves reconstruction quality substantially, achieving up to 38\% PSNR gain and approximately 28\% improvement in CLIP score compared with uniform semantic compression baselines~\cite{zhang2025multimodal}, demonstrating the effectiveness of MLLM-guided context compression.

\vspace{-1em}
\subsection{Context Management}


Context management maintains and updates wireless context over time to preserve effectiveness under dynamic environments and task requirements. Since static context rapidly becomes outdated due to mobility and environmental changes, continuous maintenance is essential. In~\cite{wang2025radio}, a radio environment knowledge pool (REKP) is proposed to support structured context updates via both new observations and task-driven feedback, while preventing uncontrolled context growth through knowledge sorting and replacement. Numerical results demonstrate that the continuously updated REKP achieves more than 1~dB reduction in path-loss prediction error at the 80\% cumulative distribution function point compared with learning-based baselines, highlighting the effectiveness of structured context management in sustaining prediction accuracy under dynamic propagation conditions~\cite{wang2025radio}.











\vspace{-1em}
\subsection{Lessons Learned}

From the above analysis, several key lessons emerge for improving wireless and edge intelligence through contextual information. First, context should be treated as a first-class system resource, as system performance is increasingly determined by how effectively inference-time context is selected and utilized rather than by model capacity alone. Moreover, context relevance is inherently dynamic and task-dependent, varying across system states and time scales, which makes static or one-shot context provisioning insufficient. In addition, context handling is intrinsically distributed, since context is generated, processed, and consumed across multiple network entities, requiring coordinated system-level management beyond isolated model optimization. Finally, well-engineered context directly translates into intelligence gains: delivering task-relevant context to the right agent at the right time improves prediction accuracy, generation stability, and decision quality under practical resource constraints. These insights highlight the need for a unified architectural approach to govern context production, exchange, and utilization as a core performance lever in agentic wireless systems.

\section{Wireless Context Communication Framework: Architecture and Validation}

\begin{figure*}[htp]
    \centering
    \includegraphics[width= 0.90\linewidth]{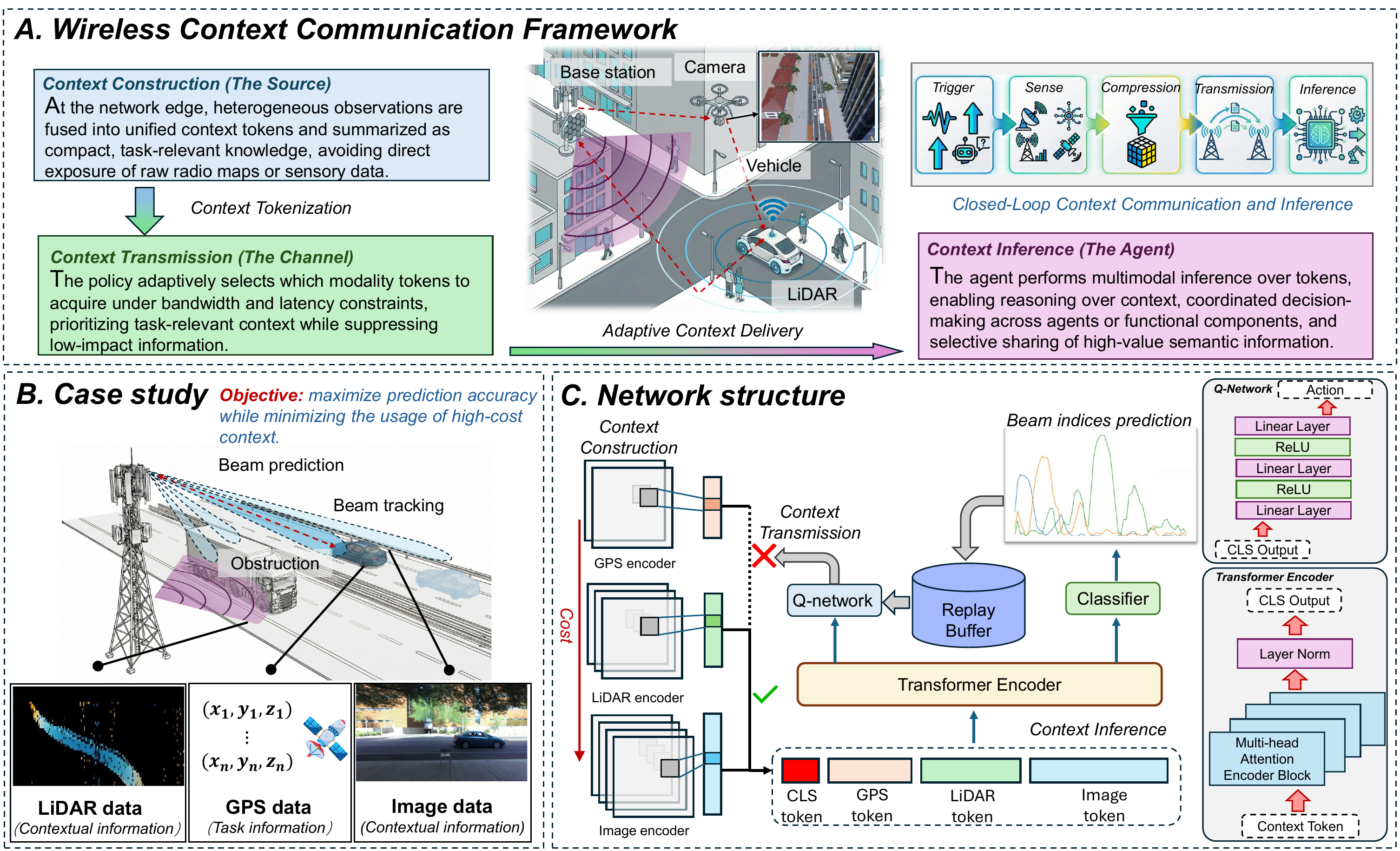}
    \caption{WCCF and ISAC-enabled beam prediction case study. 
(A) WCCF architecture with context construction, transmission, and inference, enabling closed-loop adaptive context engineering. 
(B) ISAC-enabled V2I beam prediction scenario with the objective of maximizing prediction accuracy while minimizing high-cost context usage. 
(C) Network structure of WCCF, including modality-specific encoders, a multimodal Transformer, and an RL-based policy.
}
    \label{fig:fig2}
\end{figure*}


This section presents the Wireless Context Communication Framework (WCCF), together with its architectural design and experimental validation, to enable inference-time context orchestration for agentic wireless intelligence under stringent computation, latency, and bandwidth constraints. 

\vspace{-0.5em}
\subsection{Framework Architecture}

As illustrated in Fig.~\ref{fig:fig2} Part A, WCCF consists of three tightly coupled planes.

\begin{enumerate}

\item \textbf{Context Construction (The Source):}
The Context Construction module at the network edge fuses heterogeneous observations into retrievable context tokens. Modality-specific encoders extract visual, geometric, and physical features, which are integrated via cross-attention or late fusion into a unified embedding space~\cite{xu2025integrating}. In macrodiversity or cell-free deployments, context construction is distributed across access points. Each base station extracts multi-timescale local descriptors, such as long-term shadowing or mobility patterns, which are then fused into unified tokens. This collaborative approach enables the agent to select the optimal transmission path by reasoning over the global context.


\item \textbf{Context Transmission (The Channel):}
Rather than streaming raw data, WCCF focuses on adaptive context selection to complement essential task data, such as vehicle position or trajectory. Additional data, including image and LiDAR features, are treated as contextual data and acquired only when expected to improve decision quality. Under bandwidth and latency constraints, an RL-based policy determines which modality tokens to retrieve, prioritizing task-relevant context while suppressing low-impact information~\cite{mei2025survey}. This on-demand mechanism delivers only decision-critical context on top of basic task inputs, reducing redundant semantic sensing and avoiding performance degradation from excessive inputs.

\item \textbf{Context Inference (The Agent):}
The agent employs a gated multimodal Transformer to process arbitrary subsets of retrieved tokens. Masked attention and learnable missing tokens enable robust inference in the absence of modalities~\cite{liu2025context}. This design enables reasoning over accumulated context, supports coordinated decision-making across multiple agents or functional components, and allows selective sharing of high-value semantic information when needed.

\end{enumerate}

To concretize WCCF, we implement a multimodal Transformer-based agent that represents heterogeneous wireless context as unified tokens and adaptively acquires them via reinforcement learning, integrating three core mechanisms corresponding to context acquisition, transmission, and inference, as illustrated in Fig.~\ref{fig:fig2} Part C.

\begin{itemize}

\item \textbf{Multimodal Context Tokenization:} 
Each modality is first processed by a dedicated encoder to extract modality-specific features, which are then projected into a shared embedding space to form unified context tokens. This tokenization process converts heterogeneous inputs, including GPS, images, and LiDAR Scene Coordinate Regression (SCR), into a consistent sequence representation that can be jointly consumed by the Transformer.

\item \textbf{RL-Based Modality Selection:} 
An RL policy dynamically determines which context tokens to retrieve at each time step. The agent decides whether to rely solely on task information or to acquire additional contextual descriptors, choosing between lower-cost LiDAR features and higher-cost image cues to assist task execution, explicitly balancing performance gains against communication and processing overhead.

\item \textbf{Masked Multimodal Inference:} 
To support operation under partial context, the Transformer is trained with modality masking, where unavailable modalities are replaced by mask tokens. This simple strategy enables the model to handle arbitrary modality subsets at inference time, preserving architectural consistency while adapting its predictions to the available context.

\end{itemize}

\subsection{Runtime Context Lifecycle}
\label{subsec:runtime_lifecycle}

Beyond the static decomposition of Fig.~\ref{fig:fig2} Part A, WCCF specifies how contextual modalities are handled during runtime decision making under limited inference-time capacity.

\textit{Who triggers context acquisition?}
Context acquisition is controlled by the agent policy rather than a fixed sensing schedule. At each step, the agent decides whether to retrieve additional contextual modalities to support the downstream task.

\textit{When is additional context retrieved?}
Additional context is acquired only when its expected utility justifies the associated sensing and encoding cost. In the case study, this trade-off is reflected in the reinforcement learning reward that jointly considers prediction performance and modality-dependent acquisition overhead.

\textit{How is context represented?}
Heterogeneous modalities, including GPS, image, and LiDAR observations, are converted into unified context tokens through modality-specific encoders before multimodal inference.

\textit{How is partial context handled?}
The inference model operates across arbitrary modality subsets via masked multimodal inference, enabling robust performance even when only partial context is available.

\subsection{Case Study}

We evaluate WCCF on an ISAC-enabled beam prediction task using the DeepSense 6G V2I dataset (Scenario~9)\footnote{https://www.deepsense6g.net}, where a base station continuously selects transmit--receive beam pairs for a moving vehicle under dynamic blockage. In this setting, vehicle location serves as task information for beam prediction, while image and LiDAR observations provide optional contextual information. This scenario naturally exhibits heterogeneous context costs: low-dimensional location information is inexpensive, whereas acquiring image and LiDAR context incurs substantial computation overhead.

Within WCCF, GPS-based location is treated as task data and is always available, while \textbf{Image} and \textbf{LiDAR} are considered contextual data acquired on demand. The RL agent optimizes a reward defined as a binary top-3 beam prediction indicator at the next time step (reward $=1$ if the correct beam is within top-3, and $0$ otherwise), penalized by modality-dependent acquisition costs reflecting encoder complexity. Specifically, the cost coefficients for GPS, Image, and LiDAR are set to $0.01$, $0.1$, and $0.9$, respectively. This formulation encourages the agent to maximize prediction accuracy while minimizing reliance on expensive semantic sensing.

\subsubsection{Performance Evaluation}

We evaluate WCCF under four inference settings: \textbf{Only\_GPS}, where only low-cost task information is available; \textbf{Missing\_image} and \textbf{Missing\_LiDAR}, where one contextual modality is absent; and \textbf{Full\_observation}, where all modalities are continuously accessible. These fixed configurations serve as baselines with different context availability, against which we compare the proposed RL-driven adaptive context selection policy.

\begin{figure}[tbp]
\centering
\includegraphics[width=0.75\linewidth]{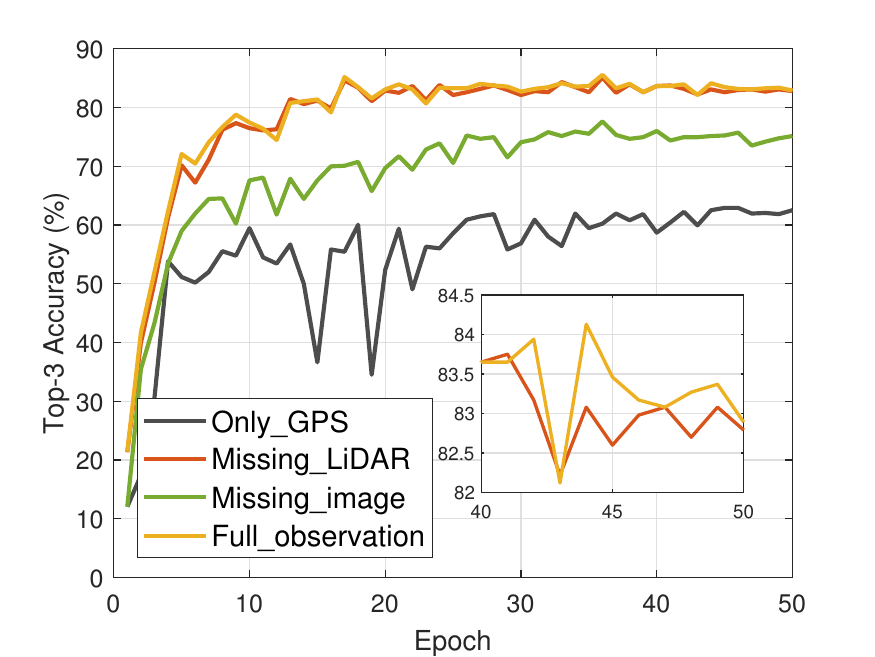} 
  \caption{Top-3 beam prediction accuracy under different context configurations}
\label{fig:ex1}
\end{figure}

\begin{figure}[tbp]
\centering
\includegraphics[width=0.75\linewidth]{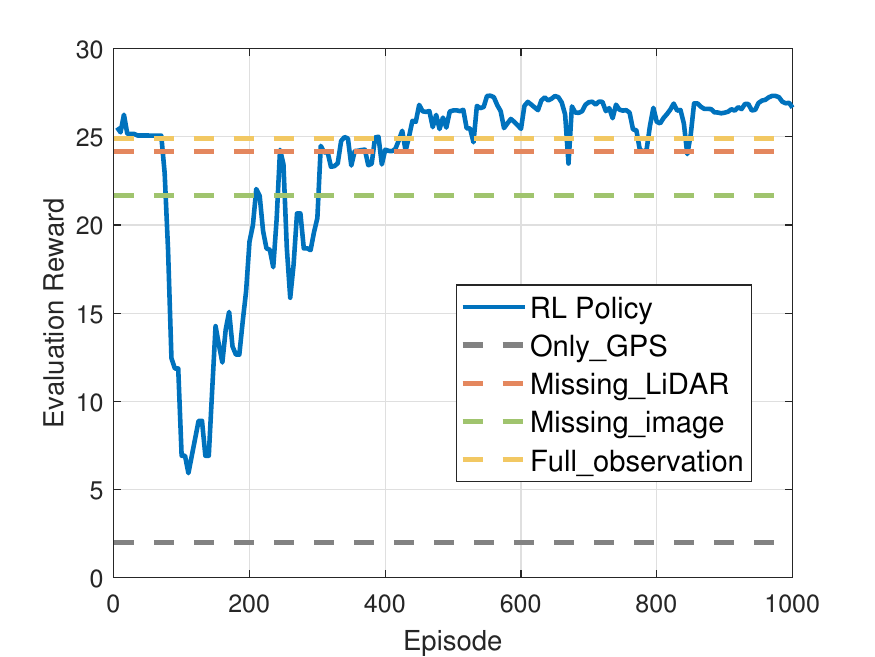} 
  \caption{Evaluation reward of the RL-driven WCCF policy during training, compared with fixed context baselines}
\label{fig:ex2}
\end{figure}

Fig.~\ref{fig:ex1} reports the top-3 beam prediction accuracy under different context configurations, while Fig.~\ref{fig:ex2} shows the corresponding evaluation reward that jointly accounts for prediction accuracy and context acquisition cost. Using only positional task data yields approximately 60-63\% accuracy, highlighting the insufficiency of task information under dynamic blockage. Introducing a single contextual modality improves performance to around 74-76\%, and continuous full observation further increases accuracy to about 83-85\%. This indicates that wireless context contributes roughly 20 percentage points over GPS-only inference, whereas the marginal gain of full multimodal sensing over two modality context is limited to approximately 3-5 percentage points, revealing diminishing returns from indiscriminate context acquisition.

More importantly, the RL-driven WCCF policy rapidly converges to a stable strategy and achieves evaluation rewards comparable to the full-observation baseline while avoiding persistent access to high-cost semantic inputs. Compared with fixed single-modality configurations, the adaptive policy improves the overall reward by approximately 10-25\% and operates within 2-3\% of full-observation accuracy. These results demonstrate that wireless context is highly state-dependent, and that near-optimal beam prediction can be achieved through adaptive context orchestration with substantially reduced inference burden, validating the central premise of wireless context engineering: effective edge intelligence arises from judicious context utilization rather than maximal information access.

\section{Conclusion}

In this paper, we have explored wireless context engineering as a foundational paradigm for enabling efficient edge general intelligence under limited inference-time capacity. We systematically characterized wireless context across multiple system layers and introduced five key engineering dimensions. Building on these principles, we proposed a WCCF to support adaptive multimodal context engineering under finite bandwidth and inference-cost constraints. Through an ISAC–enabled V2I beam prediction case study, we have demonstrated that dynamic selection of wireless context achieves an optimized trade-off between prediction accuracy and sensing cost, validating the effectiveness of principled context engineering over indiscriminate multimodal acquisition.



\bibliography{Ref}

\end{document}